\documentclass{revtex4}
\usepackage{graphicx}
\usepackage{natbib}
\usepackage{epsfig}

\begin{document}

\title{Generalized statistical mechanics for superstatistical systems}

\author{Christian Beck}

\affiliation{School of Mathematical Sciences, Queen Mary University of London,
Mile End Road, London E1 4NS, UK}

\begin{abstract}
Mesoscopic systems in a slowly fluctuating environment are often well described
by superstatistical models.
We develop a generalized statistical mechanics formalism for
superstatistical systems, by mapping the superstatistical
complex system onto a system of ordinary statistical mechanics with
modified energy levels.
We also briefly review recent examples of applications
of the superstatistics concept for three very different
subject areas, namely train delay statistics,
turbulent tracer dynamics, and cancer survival statistics.
\end{abstract}

\maketitle

\section{Introduction}
Nonlinear dynamical processes often create a fluctuating environment
for a given mesoscopic system \citep{beck-cohen}.
 This leads to mixing of the
mesoscopic dynamics and that of the environment. If there is sufficient time scale separation,
then very often superstatistical models yield a good effective
description.
The superstatistics concept has established itself as a powerful tool
to describe quite general classes of complex systems
\citep{beck-cohen,unicla,touchette,supergen,souza,chavanis,vignat,Plastino-x,jizba,
wilk,prl01,frank,hase,celia,abul-again,straeten-recent}. The basic idea is to characterize the
complex system under consideration by a superposition of several
statistics on different time scales, for example one corresponding
to ordinary statistical mechanics (on a mesoscopic level modelled by a
Langevin equation) and the other one corresponding to a slowly
varying inverse temperature field $\beta (\vec{x},t)$ or some other
relevant parameter.

There may be either spatial or temporal
variations of the environment. The environment is represented by a suitable parameter
entering the stochastic differential equation describing
the mesoscopic system. The superstatistics concept can be applied in
quite a general way, and a couple of interesting applications
for a
variety of complex systems have been pointed out recently
\citep{daniels,maya,reynolds,RMT,abul-magd3,porpo,rapisarda,kantz,cosmic,eco,straetennew,eco2}.
Essential for this approach is the existence of sufficient time
scale separation so that the system has enough time to relax
to a local equilibrium state and stay within it for some time.
%There is an intensive parameter $\beta$ that fluctuates
%on a much larger time scale than the typical relaxation time of
%the local dynamics. In a thermodynamic setting, $\beta$ can be
%interpreted as a local inverse temperature of the system, but
%much broader interpretations are possible.

The stationary distributions of superstatistical systems,
obtained by averaging over all $\beta$, typically exhibit
non-Gaussian behavior with fat tails, which can be a power law, or
a stretched exponential, or other functional forms as well
\citep{touchette}. In general, the superstatistical parameter
$\beta$ need not to be an inverse temperature but can be an
effective parameter in a stochastic differential equation, a
volatility in finance, or just a local variance parameter
extracted from some experimental time series. There are interesting applications
in hydrodynamic
turbulence \citep{prl,beck03,reynolds,unicla}, for defect
turbulence \citep{daniels}, for cosmic rays \citep{cosmic}
and other scattering processes in high energy physics \citep{wilknew, superscatter}, solar
flares \citep{maya}, share price fluctuations
\citep{bouchard,ausloos,eco,straetennew}, random matrix theory
\citep{RMT,abul-magd2,abul-magd3}, random networks
\citep{abe-thurner}, multiplicative-noise stochastic processes
\citep{queiros}, wind velocity fluctuations \citep{rapisarda, kantz},
hydro-climatic fluctuations \citep{porpo}, the statistics of train
departure delays \citep{briggs} and survival statistics of
cancer patients \citep{chen}.
Maximum entropy principles can be generalized in a suitable
way to yield the relevant probability distributions
that characterize the various important universality classes
in superstatistics \citep{souza, abc, crooks,
naudts, straeten}.

In this paper we shall develop a new theoretical approach to superstatistics,
by formally mapping the superstatistical system onto a system of ordinary
statistical mechanics where the energy levels are modified in a suitable
way. This approach yields a new interesting theoretical tool to further develop
the generalized statistical mechanics of superstatistical complex systems,
and is described in detail in section 3.
We also briefly review some recent examples of applications
of superstatistical techniques. Our three examples,
all from very different subject areas, are train delays
on the British railway network, velocity signals in hydrodynamic
turbulence, and the survival statistics of cancer patients.

\section{Reminder: What is superstatistics?}\label{sec1}

The concept is best illustrated by starting with a particular example
of superstatistics, in fact the one that was considered first in
\citep{wilk, prl01}.
Consider
the following well-known formula:

\begin{equation}
\int_0^\infty d \beta f(\beta) e^{-\beta E}
=\frac{1}{(1+(q-1)\beta_0 E)^{1/(q-1)}} \label{111}
\end{equation}
where
\begin{equation}
f (\beta) = \frac{1}{\Gamma \left( \frac{1}{q-1} \right)} \left\{
\frac{1}{(q-1)\beta_0}\right\}^{\frac{1}{q-1}}
\beta^{\frac{1}{q-1}-1} \exp\left\{-\frac{\beta}{(q-1)\beta_0}
\right\} \label{fluc}
\end{equation}
is the $\chi^2$ (or $\Gamma$) probability distribution and $\beta_0$ and $q$
are parameters ($q>1$).

We see that averaged {\em ordinary} Boltzmann factors $e^{-\beta E}$ with
$\chi^2$-distributed $\beta$ yield an {\em effective} Boltzmann
factor of $q$-exponential form,
given by the right-hand side of eq.~(\ref{111}). The physical interpretation is
that nonequilibrium systems with temperature fluctuations
give rise to an effective description in terms of more general
Boltzmann factors.
In
\citep{wilk,prl01} the $\chi^2$-distribution was advocated for
$f(\beta)$, because at that time the aim was to better
understand $q$-statistics \cite{tsallis} from a dynamical point of view.
% see also \cite{?} as an
%example that similar ideas can be found in much older papers.
General $f(\beta)$
were then suggested in \citep{beck-cohen}. In that paper also the name `superstatistics'
was created. This name was simply an abbreviation for the fact that there is a {\em superposition}
of two (or several) statistics. In no way this name wants to indicate
that this type of statistics is `superior' to others.

One can also construct dynamical realizations of superstatistics in terms of Langevin
equations with parameters that fluctuate on large time scales \citep{prl01}.
These local Langevin equations decribe the mesoscopic system under
consideration. The situation is sketched in Fig.~1.
\begin{figure}
\includegraphics[width=8cm]{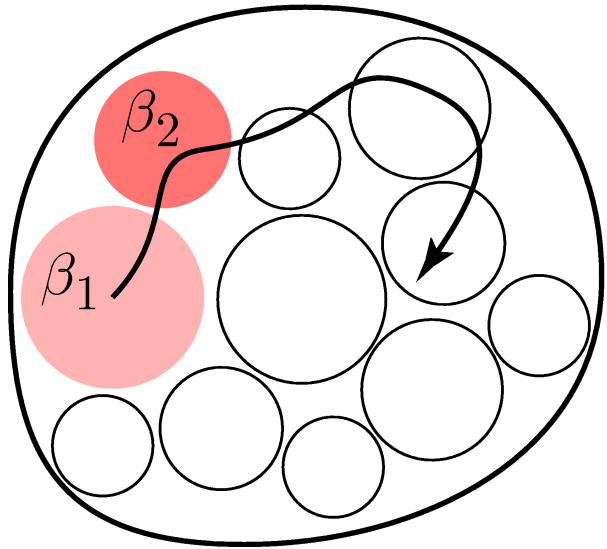}
\caption{A spatially inhomogeneous situation of mesoscopic systems (sketched
as circles) embedded
into a fluctuation environment with different inverse temperatures $\beta_i$.
A Brownian particle moves through the different regions with different inverse
temperatures.}
\end{figure}
The simplest example would be locally a linear Langevin equation
\begin{equation}
 \dot{v}=-\gamma v + \sigma L(t)
\end{equation}
 with slowly
 fluctuating parameters $\gamma , \sigma$.
Here $L(t)$ denotes Gaussian white noise.
This describes the velocity $v$ of
a Brownian particle that moves through spatial
'cells' with different local $\beta:=\gamma /(2\sigma^2)$ in each
cell (a nonequilibrium situation).
%nonequilibrium system with spatio-temporal fluctuations of
%inverse temeperature $\beta \sim \gamma /\sigma^2$ on relatively
%long time scale)
If some probability distribution $f(\beta)$ of the inverse temperature $\beta$ for the various
cells is given,
then the conditional probability given some fixed $\beta$
in a given cell is Gaussian,
$p(v|\beta) \sim e^{-\frac{1}{2} \beta v^2}$, the joint probability
is $p(v, \beta)=f(\beta) p(v|\beta)$ and the marginal probability
is $p(v)=\int_0^\infty f(\beta) p(v|\beta) d\beta$. Integration
over $\beta$ yields effectively Boltzmann factors that are more
general than Gaussian distributions, which depend on the specific
properties of $f(\beta)$. If there are only finitely many cells,
then the integral is understood
to approximate the average over a large number of cells.

The principal idea of superstatistics is to generalize this example to much
broader systems. For example, $\beta$ need not be an inverse
temperature but can in principle be any intensive parameter. Most
importantly, one can generalize to {\em general probability
densities $f(\beta)$} and {\em general Hamiltonians}. In all
cases one obtains a superposition of two different statistics:
that of $\beta$ and that of ordinary statistical mechanics.
Superstatistics hence describes complex {nonequilibrium systems}
with { spatio-temporal fluctuations of an intensive parameter} on
a large scale. The {\em effective} Boltzmann factors $B(E)$ for
such systems are given by

\begin{equation}
B(E)=\int_0^\infty f(\beta) e^{-\beta E} d\beta .
\end{equation}
%Here $f(\beta)$ is the probability distribution of $\beta$. Many
%results can be proved for {\em general $f(\beta)$}.

%\item $f(\beta)$ =$\chi^2$-distribution $\Rightarrow$ Tsallis
%statistics obtained as a special case of general superstatistics.
%Marginal distributions given by $B(E)\sim e_q^{-\tilde{\beta} E}$.

%\item Tsallis statistics has dynamical realization in terms of
%Langevin equation with fluctuating parameters (C.B. Phys. Rev.
%Lett. 2001)

%\newpage

%\begin{itemize}
%\item
%For sharply peaked $f(\beta)$ all superstatistics approach
%Tsallis statistics in a {\color{red} universal} way.

%\item
%\begin{displaymath}
%q=\frac{\langle \beta^2 \rangle}{\langle \beta \rangle^2}
%\end{displaymath}

%(no fluctuations: $\Rightarrow q \to 1$).

%\item
%$\beta$: can quite generally be some intensive variable or some
%system parameter describing environment

%\end{itemize}

A lot of research has been done in this direction in recent years.
If there is locally Gaussian behaviour, then the
theory of superstatistics is clearly related to the theory of Gaussian
scale mixtures. More generally 
one can prove a superstatistical generalization of fluctuation
theorems \citep{supergen},
develop a variational principle for the large-energy
asymptotics of general superstatistics \citep{touchette},
proceed to generalized entropies for general
superstatistics \citep{souza,abc, straeten},
let the $q$-values in eq.~(\ref{111}) fluctuate as well
\citep{vignat}, and prove
superstatistical versions of a Central Limit
Theorem \citep{Plastino-x}.
There are also relations with
fractional reaction equations \citep{hau},
random matrix theory \citep{RMT, abul-magd2, abul-magd3},
networks \citep{abe-thurner}, and
path integrals \citep{jizba}. Very useful for practical applications is
a superstatistical approach to
time series analysis \citep{unicla,kantz, straetennew}.
Applications have been pointed out for
3d
hydrodynamic turbulence \citep{beck03,reynolds,unicla,prl},
wind velocity
fluctuations \citep{rapisarda,kantz},
finance
and economics \citep{ausloos, eco2, heston, bouchard},
blinking quantum dots \citep{grigolini},
cosmic ray statistics \citep{cosmic} and quite generally
scattering processes in particle
physics \citep{wilknew,superscatter}. The concept has also been
useful to analyze
hydroclimatic fluctuations \citep{porpo} as well as
the statistics of train delays on the British railway network \citep{briggs}.
There are also
medical applications \citep{chen}.

\section{Mapping superstatistics onto conventional statistical mechanics}

Consider a system of ordinary statistical mechanics with energy levels $E_i$
of microstate $i$. We are looking at a canonical ensemble and {\em a priori} the inverse
temperature $\beta$ is fixed. Now look at identical copies of the system but with different
temperatures $\beta_j$ in each spatial cell $j$, at a given snapshot of time. This
is a nonequilibrium situation.

Let $\beta_0=\int_0^\infty f(\beta) \beta d\beta$ be the average inverse temperature.
We may formally consider a super-Hamiltonian describing the entire system which in the different
spatial cells has effective energy levels $\tilde{E}_i$, by writing
\begin{equation}
\beta_0 \tilde{E}_i = \beta_j E_i.
\end{equation}
Apparently this means the super-Hamiltonian has energy levels $\tilde{E}_i^{(j)}$
given by
\begin{equation}
\tilde{E}_i^{(j)}= \frac{\beta_j}{\beta_0}E_i
\end{equation}
in cell $j$.

Since ordinary statistical mechanics is valid for {\em arbitrary} energy levels,
in particular also for the $\tilde{E}_i^{(j)}$, we may now do
ordinary statistical mechanics for the super-Hamiltonian and introduce the partition
function $Z(\beta_0)$ of the entire system as
\begin{eqnarray}
Z(\beta_0)&=&\sum_{j,i}e^{-\beta_0\tilde{E}_i^{(j)}} \\
\, &= & \sum_{j,i} e^{-\beta_j E_i} \\
\, &\approx & \int_0^\infty f(\beta) e^{-\beta E_i}
 d\beta ,
\end{eqnarray}
where in the last step the sum over $j$ is approximated by an integral.
Since $Z(\beta_0)=\sum_{j,i}e^{-\beta_0 \tilde{E}_i^{(j)}}$ is an ordinary
partition function (though with exotic, locally modified, energy levels), it is now possible to
do ordinary statistical mechanics for this superstatistical nonequilibrium system, with all
the known formulas.

We regard the free energy of the superstatistical system as a function of the
mean inverse temperature $\beta_0$ and define it as
\begin{equation}
F(\beta_0)=-\frac{1}{\beta_0} \log Z(\beta_0).
\end{equation}
In the statistical mechanics formalism it is often convenient to work with
the function $\Psi (\beta_0):= \beta_0 F(\beta_0)$. Defining $kT_0=\frac{1}{\beta_0}$
one has
\begin{eqnarray}
F&=&U-T_0S \\
\Psi&=&\beta_0U-S
\end{eqnarray}
But one has to be careful here what the meaning of the symbols $U$ and $S$ is:
$U$ is now the mean energy of the energy levels $\tilde{E}_i$, rather than $E_i$,
and indeed this means that $U$ is a global mean energy corresponding to the entire superstatistical system
consisting of many cells. One has
\begin{eqnarray}
U&=&\sum_{j,i} p_i^{(j)}\tilde{E}_i^{(j)}\\
\,&=&\sum_{j,i} p_i^{(j)} \frac{\beta_j}{\beta_0} E_i \\
\,&\approx& \frac{1}{\beta_0} \int_0^\infty d\beta f(\beta) \beta U_{loc}(\beta),
\end{eqnarray}
where $U_{loc}(\beta)=\sum p_iE_i$ is the local internal energy in a cell of inverse temperature $\beta$.
The entropy is still given by the Boltzmann-Gibbs-Shannon form, but formed with
the exotic energy levels $\tilde{E}_i^{(j)}$:
\begin{eqnarray}
S &=& -k\sum_{j,i} p_i^{(j)}\log p_i^{(j)} \\
\, & =& -k\sum_{j,i} \frac{1}{Z(\beta_0)}e^{-\beta_0\tilde{E}_i^{(j)}}( -\log Z(\beta_0)
-\beta_0 \tilde{E}_i^{(j)}) \\
\, &=& \beta_0 U +\log Z(\beta_0)
\end{eqnarray}
In this way we have formally mapped the superstatistical nonequilibrium system
onto an (exotic) equilibrium system of ordinary statistical mechanics with average
inverse temperature $\beta_0$ and a new type of Hamiltonian, corresponding to the
energy levels $\tilde{E}_i^{(j)}$. We should remark that the above
idea of mapping superstatistics onto the
statistical mechanics of an exotic Hamiltonian is completely new
and different from previous attempts of developing a generalized statistical
mechanics for superstatistical systems \citep{souza, abc, crooks, straetennew}.

\section{Possible superstatistical distributions $f(\beta)$}

The distribution $f(\beta)$ is determined by the dynamical large-scale structure of the
complex system under consideration. There have been attempts to derive
the specific form of $f(\beta)$ relevant for a given complex system
with given constraints from a generalized maximum entropy principle.
We don't elucidate this further here but refer to \citep{straeten} and
references therein for further details. Actually, what we want to
do here is to proceed
to practical applications. The relevant question is what type of $f(\beta)$
are typically seen for experimental data as generated by a generic complex system.

There seem to be three different superstatistics that are
of utmost importance \citep{unicla}. These are
(a) $\chi^2$-superstatistics ($=$ Tsallis statistics),
(b) inverse $\chi^2$-superstatistics, and
(c) lognormal superstatistics.

In case (a), $f(\beta)$ is given by the Gamma distribution
\begin{equation}
f(\beta )=\frac 1{\Gamma (\frac n2)}\left( \frac n{2\beta
_0}\right) ^{n/2}\beta ^{n/2-1}e^{-\frac{n\beta }{2\beta _0}},
\label{chi2}
\end{equation}
where again $\beta_0$ is the average of $\beta$. This generates generalized
Boltzmann factors $B(E)$ that decay with a power law. $n$ is a
parameter characterizing the number of degrees of freedom.

In case (b), $f(\beta)$ is given by
\begin{equation}
f(\beta )=\frac{\beta _0}{\Gamma (\frac n2)}\left( \frac{n\beta
_0}2\right) ^{n/2}\beta ^{-n/2-2}e^{-\frac{n\beta _0}{2\beta }}.
\label{chi2inv}
\end{equation}
In this case the generalized Boltzmann factors  $B(E)\sim \int
f(\beta) e^{-\beta E}$ decay as $e^{-\tilde{\beta}\sqrt{E}}$
for large $E$.

%\pause Again this superstatistics is {\color{red} universal}:
%details of the $\xi_j$ are irrelevant.

Finally, in case (c) $f(\beta)$ is given by the lognormal
distribution
\begin{equation}
f(\beta )=\frac{1}{\sqrt{2\pi}s\beta} \exp \left\{ \frac{-(\ln
\frac{\beta}{\mu})^2}{2s^2}\right\}, \label{logno}
\end{equation}
where $\mu$ and $s$ are suitable parameters. In the remaining sections, we
briefly describe one example for each of these three different cases.
%\pause

%Again this is a {\color{red} universal} result, details of the
%$\xi_j$ are irrelevant.

\section{Train departure delays}\label{sec3}

Traffic delays on the British railway network are reasonably well described
by $\chi^2$-superstatistics. The probability density of observed train departure
delays of length $t$ has been analyzed in detail in \citep{briggs}.
Millions of departure times were automatically stored and evaluated.
%Clearly,
%at the various stations there are sometimes train departure delays
%of length $t$.
The 0th-order theoretical model for the waiting time $t$ is
a Poisson process which predicts that the waiting time
distribution until the train finally departs is $
P(t|\beta)=\beta e^{-\beta t}$, where $\beta$ is some parameter.
But this does not agree with the actually observed data
\cite{briggs}. A much better fit is given by a $q$-exponential,
see Fig.~2.

\begin{figure}
\includegraphics[width=8cm]{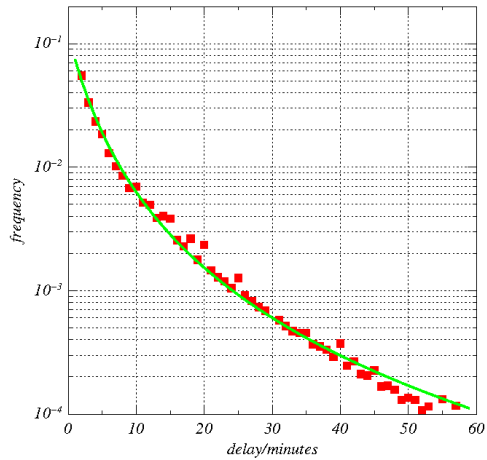}
\caption{Observed departure delay statistics on the British
railway network (data from \citep{briggs}). The solid line is a
q-exponential fit as given by eq.~(\ref{555}).}
\end{figure}

What may cause this power law that fits the data? The idea is that
there are fluctuations in the parameter $\beta$ as well. These
fluctuations describe large-scale temporal or spatial variations
of the British rail network environment, which take place on a much
larger time scale than the actual train departures.
$\beta$-fluctuations are e.g. produced at
the begin of the holiday season with lots of passengers, or if there
are problems with the track or
bad weather conditions.
Also there can be extreme events such as derailments, industrial action, terror
alerts, etc.
The observed long-term distribution of train delays is then a
mixture of exponential distributions where the parameter $\beta$
fluctuates:
\begin{equation}
p(t)=\int_0^\infty f(\beta) p(t|\beta) d\beta = \int_0^\infty
f(\beta) \beta e^{-\beta t}. \label{9}
\end{equation}
For a $\chi^2$-distributed $\beta$ with $n$ degrees of freedom
one obtains
\begin{equation}
%p(t) \sim \frac{1}{\left( 1+b(q-1)t\right)^{\frac{1}{q-1}}}
p(t) =C \cdot {\left( 1+b(q-1)t\right)^{\frac{1}{1-q}}} \label{555}
\end{equation}
where $q=1+{2}/(n+2)$, $b= {2}\beta_0 /({2-q})$ and $C$ is a
normalization constant.
%\begin{equation}
%q=1+\frac{2}{n+2}
%\end{equation}
%and
%\begin{equation}
%b= \frac{2}{2-q} \beta_0.
%\end{equation}
%The model discussed in \cite{briggs} generates $q$-exponential
%distributions of train delays by a simple mechanism, namely a
%$\chi^2$-distributed parameter $\beta$ of the local Poisson
%process. This is an example for $\chi^2$ superstatistics.

\section{Turbulent flows}
%
%Navier-Stokes equation: $ \dot{v}=-(v\nabla)v-\frac{1}{2}\sigma^2
%\Delta v +F$

%Turbulence $=$ spatio-temporal chaotic state of the Navier-Stokes
%equation.

%\pause
Various aspects of hydrodynamic turbulence are
quite well described by lognormal
superstatistics. In this case the
mesoscopic local dynamics corresponds to
a single tracer particle that is advected by the turbulent flow.
The environment of the tracer particle changes.
For a while it will see regions of strong turbulent
activity, then it will move on to calmer regions, just to continue in yet
another region of strong activity, and so on. This is a
superstatistical dynamics similar to Fig.~1.
In 'Lagrangian turbulence' one is interested
in the statistics of velocity differences
$\vec{u}(t):=\vec{v}(t+\tau) -\vec{v} (t)$ of the particle on a
small time scale $\tau$. For $\tau \to 0$ this velocity
difference becomes the local acceleration $\vec{a}
(t)=\vec{u}(t)/\tau$. A superstatistical Lagrangian model for
{3-d} velocity differences of the tracer particle has been
developed in \citep{prl}. The mesoscopic dynamics is a superstatistical
Langevin equation of the form

\begin{equation}
\dot{\vec{u}}=-\gamma \vec{u}+B \vec{n} \times \vec{u}+\sigma
\vec{L}(t). \label{2}
\end{equation}
%This example corresponds to a choice of matrices $\Gamma$ and
%$\Sigma$ in eq.~(\ref{1}) where $\Sigma= \sigma \cdot {\bf 1}$ is
%proportional to the identity matrix, whereas $\Gamma$ also
%contains nondiagonal entries if $B\neq 0$.
Here $\gamma$ and $B$ are constants. The term
proportional to $B$ introduces some rotational movement of the
particle, simulating vortices in the flow. The noise strength
$\sigma$  and the unit vector $\vec{n}$ evolve stochastically on
a large time scale $T_\sigma$ and $T_{\vec{n}}$, respectively.
$T_\sigma$ is of the
same order of magnitude as the integral time scale $T_L$, whereas
$\gamma^{-1}$ is of the same order of magnitude as the Kolmogorov
time scale $\tau_{\eta}$. In this model the Reynolds number
$R_\lambda$ is basically given by the time  scale ratio $T_\sigma
\gamma \sim T_L/\tau_{\eta} \sim R_\lambda
>>1$.
The
time scale $T_{\vec{n}}>> \tau_\eta$ describes the average life
time of a region of given vorticity surrounding the test particle.
Further details are described in \citep{prl}.

The parameter
$\beta$ is again defined to be
$\beta:=2\gamma/\sigma^2$, but it does {\em not} have the meaning
of a physical inverse temperature in the flow. Rather, one has
$ \beta^{-1} \sim \nu^{1/2} \langle \epsilon \rangle^{-1/2}
\epsilon$, where $\nu$ is the kinematic viscosity and $\langle
\epsilon \rangle$ is the average energy dissipation. $\epsilon$ is known to fluctuate in
turbulent flows. Kolmogorov's theory of 1961 suggests a lognormal
distribution for $\epsilon$, which leads us to lognormal
superstatistics.
% It is reasonable to assume that the
%probability density of the stochastic process $\beta(t)$ is
%approximately a lognormal distribution
%\begin{equation}
%f(\beta) = \frac{1}{\beta s \sqrt{2\pi}}\exp\left\{ \frac{-(\log
%\frac{\beta}{m})^2}{2s^2}\right\}. \label{Logno}
%\end{equation}
For very small $\tau$ the 1-d acceleration component of the
particle is given by $a_x=u_x/\tau$ and one gets out of the model
the 1-point distribution
\begin{equation}
p(a_x) = \frac{\tau}{2\pi s }\int_0^\infty d\beta \; \beta^{-1/2}
\exp\left\{ \frac{-(\log \frac{\beta}{m})^2}{2s^2}\right\}
e^{-\frac{1}{2}\beta \tau^2 a_x^2} , \label{10}
\end{equation}
which agrees very well with experimentally measured
data of the acceleration statistics (Fig.~3).

\begin{figure}
\includegraphics[width=8cm]{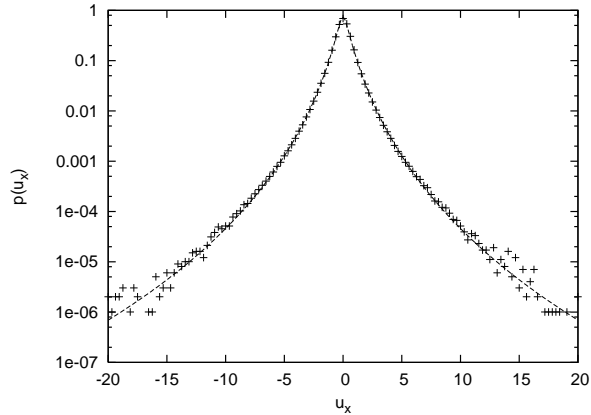}
\caption{Measured probability density of small-scale velocity
differences in Lagrangian turbulence (data from \citep{pinton})
and a fit of the form (\ref{10}) (see \citep{prl}
for more details).}
\end{figure}

The 3-d superstatistical
model of \cite{prl} predicts correlations between the three
acceleration components. An intrinsic property
of the model is that the acceleration $a_x$ in
$x$ direction is not statistically independent of the
acceleration $a_y$ in $y$-direction. We may study the ratio
$R:=p(a_x,a_y)/(p(a_x)p(a_y))$ of the joint probability
$p(a_x,a_y)$ to the product of 1-point probabilities $p(a_x)$ and $p(a_y)$.
For independent acceleration components this ratio would always
be given by $R=1$, whereas the 3-d superstatistical
model yields the prediction
\begin{equation}
R=\frac{\int_0^\infty \beta f(\beta)e^{-\frac{1}{2}\beta \tau^2
(a_x^2+a_y^2)}d\beta}{ \int_0^\infty\beta^{1/2}f(\beta
)e^{-\frac{1}{2}\beta \tau^2 a_x^2}d\beta
\int_0^\infty\beta^{1/2}f(\beta)e^{-\frac{1}{2}\beta \tau^2
a_y^2}d\beta} . \label{999}
\end{equation}
%This is a very general formula, valid for any superstatistics,
%for example also Tsallis statistics, obtained when $f(\beta)$ is
%the $\chi^2$-distribution.
The trivial result $R=1$ is obtained
only for $f(\beta)=\delta (\beta -\beta_0)$, i.e.\ no
fluctuations in the parameter $\beta$ at all. Fig.~4 shows
$R:=p(a_x,a_y)/(p(a_x)p(a_y))$ as predicted by lognormal
superstatistics:
\begin{figure}
\includegraphics[width=12cm]{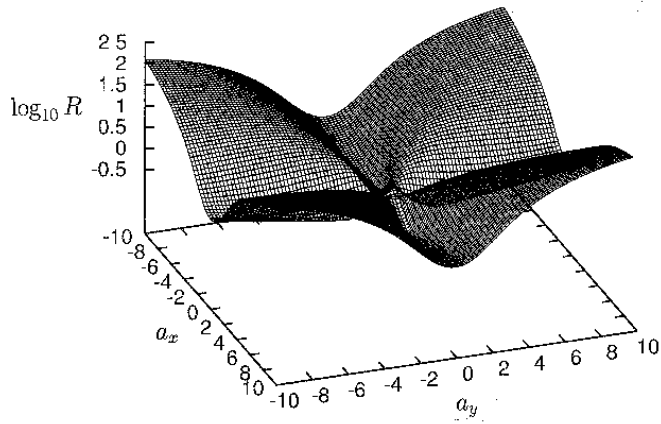}
\caption{Predicted shape of correlations between
acceleration components in a turbulent flow as given by
eq.~(\ref{999}). $f(\beta)$ is the lognormal distribution.}
\end{figure}
Experimental measurements of acceleration correlations yield very similar results
to those predicted above
\citep{boden-reynolds,prl}.

\section{Survival statistics of cancer patients}\label{sec5}
Data of the survival statistics of cancer patients can be well fitted
using models based on inverse $\chi^2$-superstatistics.
%It should be clear that once a patient is diagnosed with cancer there is usually a good
%chance of treatment, in particular in early stages, but the death occurs if metastases form.
%There are various pathways in which cancerous cells can migrate:
%Via the blood system, the lymphatic system, and so on. The
%diffusion constants for these various pathways are different. In
%this way superstatistics enters, describing different diffusion
%speeds for different pathways (see Fig.~5).
%\begin{figure}
%\includegraphics[width=8cm]{siphi-fig5.eps}
%\caption{Various pathways for the spread of cancer cells}
%\end{figure}
%But there is another important issue here: When looking at a large
%ensemble of patients then the spread of cancerous cells can be very different
%from patient to patient. For some patients the
%cancer spreads in a very aggressive way, whereas for others it is much slower and less aggressive.
%So superstatistics also arises from the fact that all patients are different.

A superstatistical model of the progression of metastasis and
the corresponding survival statistics of cancer patients has
been developed in \citep{chen}.
The final result that comes out of that
model is the following prediction for the probability
density function of survival time $t$ of a randomly chosen patient that is
diagnosed with cancer at $t=0$:
\begin{equation}
p \left( t \right) = \int _{0}^{\infty }\!{\frac
{{t}^{n-1}{\lambda}^{n}{e^{-\lambda\,t}}}{ \Gamma \left( n
\right)} \frac{\lambda_0{\left(n\lambda_0 /
2\right)}^{n/2}}{\Gamma\left(n / 2\right)} \lambda^{-n/2-2}
e^{\frac{-n\lambda_0}{2\lambda}}}{d\lambda} \label{eq6}
\end{equation}
This can also be written as
\begin{eqnarray}
p \left( t \right) &=&
\frac{(n\lambda_0)^{3n/4}}{\Gamma(n)\Gamma(n/2)} \left(
\frac{t}{2} \right)^{3n/4-1} \left[ \frac{\sqrt{2n\lambda_0t}}{n}
K_{n/2+1}\left( \sqrt{2n\lambda_0t} \right) \right. \nonumber \\
 &-& \left. K_{n/2}\left(
\sqrt{2n\lambda_0t} \right) \right], \label{eq7}
\end{eqnarray}
where $K_\nu(z)$ is the modified Bessel function. Note that this
is inverse $\chi^2$-superstati\-stics. The role of the parameter
$\beta$ is now played by the parameter $\lambda$, which in a sense
describes how aggressively the cancer propagates. This parameter
has fluctuations from patient to patient.

The above formula based on inverse $\chi^2$-superstatistics is in
good agreement with real data of the survival statistics of
breast cancer patients in the US. The superstatistical formula
fits the observed distribution very well, both in a linear and
logarithmic plot (see Fig.5).

When looking at the time scales in the above figures one should
keep in mind
that the data shown are survival distributions {\em conditioned on the fact that death occurs
from cancer}. Many patients, in particular if they are diagnosed at an early stage
and treated accordingly, will live a long
healthy life
and will die from something else than cancer. These cases are {\em not} included in the
data shown above.

\begin{figure}
\includegraphics[width=8cm]{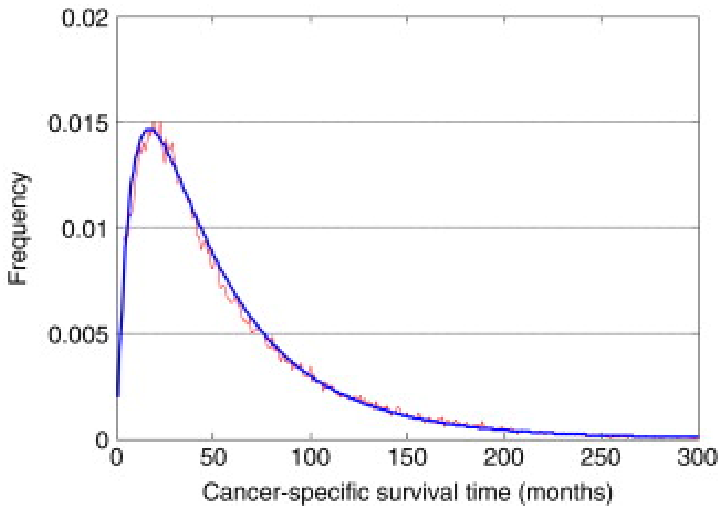}

\includegraphics[width=8cm]{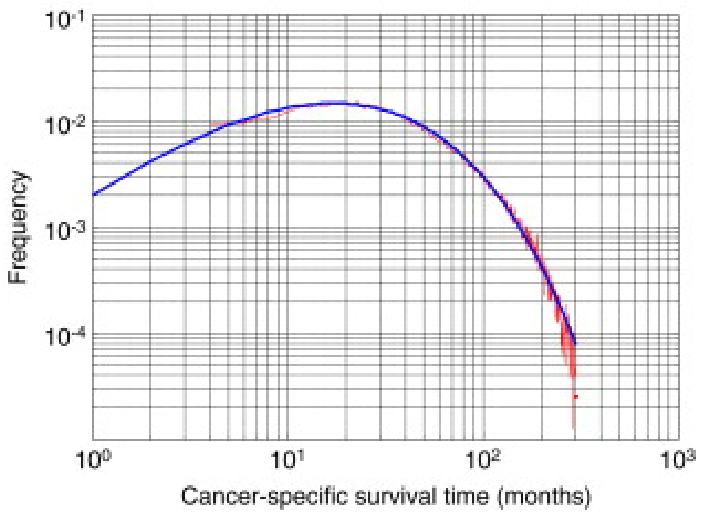}
\caption{Survival time statistics of breast cancer patients
diagnosed with cancer at $t=0$, both in a linear and double
logarithmic plot.
The solid line is the superstatistical model
prediction \citep{chen}. }
\end{figure}

\section{Conclusion and Outlook}

In this paper we have dealt with mesoscopic and other
complex systems that are embedded into
a temporally changing or spatially fluctuating environment.
If there is sufficient time scale separation, then a mixture
of different statistics (a superstatistical description) is an appropriate method
to describe these types of complex systems.
Basically this means that one mixes ordinary statistical mechanics with another
statistics of e.g. local temperature fluctuations.

In the first part of the paper we have pointed out
how  superstatistical complex systems
can be mapped onto systems of ordinary statistical mechanics.
The key point is that one deformes the effective energy levels in a suitable
way and then applies the well-known techniques of ordinary statistical
mechanics to this (exotic type of) super-Hamiltonian.

In the second part of the paper we summarized a few recent
applications of the superstatistical approach to real-world problems,
which covered quite a range of different subject areas.
We studied train
delay statistics, turbulent tracer dynamics, and survival statistics
of cancer patients.
Many other areas may
benefit from a generalized statistical mechanics formalism for
superstatistical systems as well.

In a year where there is a good chance to finally experimentally
confirm the long-awaited Higgs particle, it might be appropriate to
end this paper by mentioning that scattering processes
in high energy physics can also be well-described
by superstatistical models.
%developed in some recent papers \cite{abc,straeten}. There are
%some ambiguities which constraints should be implemented, and
%how. A very general formalism is presented in \cite{straeten},
%which contains previous work \cite{abc,crooks, naudts}  as
%special cases. The three important universality classes discussed
%above, namely $\chi^2$ superstatistics, inverse $\chi^2$
%superstatistics and lognormal superstatistics are contained as
%special cases in the formalism of \cite{straeten}. In principle,
%once a suitable generalized maximum entropy principle has been
%formulated for superstatistical systems, one can proceed to a
%generalized thermodynamic description, get a generalized equation
%of state, and so on. There is still a lot of scope of future
%research to develop the most suitable formalism. But the general
%tendency seems to be to apply maximum entropy principles and
%least biased guesses to nonequilibrium situations as well. In
%fact, Jaynes \cite{jaynes} always thought this is possible.
The experimentally observed power laws
of differential cross sections and energy spectra
at very high energies have been modelled in terms of
superstatistical generalized statistical mechanics \citep{beck00, cosmic, wilknew, superscatter}.
Superstatistical techniques have also been recently used to describe the
space-time foam in string theory \citep{kings}, and a generalized
statistical mechanics model underlying chaotic types of vacuum fluctuations
yields a Higgs mass prediction
of 154 GeV
\citep{spatio, physica-d}. It seems there is a lot of scope
for relevant contributions of generalized statistical mechanics in high energy physics
and quantum field theory.

\section*{Acknowledgements}
I am very grateful to Dr. Hugo Touchette for providing Fig.~1.


\begin{thebibliography}{16}
\providecommand{\natexlab}[1]{#1}
\expandafter\ifx\csname urlstyle\endcsname\relax
  \providecommand{\doi}[1]{doi:\discretionary{}{}{}#1}\else
  \providecommand{\doi}{doi:\discretionary{}{}{}\begingroup
  \urlstyle{rm}\Url}\fi


\bibitem[{Beck \& Cohen ({2003})}]{beck-cohen} C. Beck and E.G.D. Cohen,  Physica A {\bf 322}, 267 (2003)
\bibitem[{Beck et al. (2005)}]{unicla} C. Beck, E.G.D. Cohen, and H.L. Swinney,
Phys. Rev. E {\bf 72}, 056133 (2005)
\bibitem[{Beck \& Cohen (2004)}]{supergen} C. Beck and E.G.D. Cohen,
Physica A {\bf 344}, 393 (2004)
\bibitem{touchette} H. Touchette and C. Beck,
Phys. Rev. E {\bf 71}, 016131 (2005)
\bibitem{souza} C. Tsallis and A.M.C. Souza,
Phys. Rev. E {\bf 67}, 026106 (2003)
%\bibitem{rajagopal} A.K. Rajagopal, {\em Superstatistics -- a quantum generalization},
%cond-mat/0608679
\bibitem{jizba} P. Jizba, H. Kleinert,
Phys. Rev. E {\bf 78}, 031122 (2008)
\bibitem{vignat} C. Vignat, A. Plastino and A.R. Plastino,
cond-mat/0505580
\bibitem{Plastino-x} C. Vignat, A. Plastino, arXiv 0706.0151
\bibitem{chavanis} P.-H. Chavanis,
Physica A {\bf 359}, 177 (2006)
\bibitem{wilk} G. Wilk and Z. Wlodarczyk,
Phys. Rev. Lett. {\bf 84}, 2770 (2000)
\bibitem{prl01} C. Beck,
Phys. Rev. Lett. {\bf 87}, 180601 (2001)
%\bibitem{sattin} F. Sattin, {\em Superstatistics from a different viewpoint},
%Physica A {\bf 338}, 437 (2004)
\bibitem{daniels} K.~E. Daniels, C. Beck, and E. Bodenschatz,
Physica D {\bf 193}, 208 (2004)
%%10
\bibitem{cosmic} C. Beck,
Physica A {\bf 331}, 173 (2004)

\bibitem{maya} M. Baiesi, M. Paczuski and A.L. Stella,
Phys. Rev. Lett. {\bf 96}, 051103 (2006)
\bibitem{eco} Y. Ohtaki and H.H. Hasegawa,
cond-mat/0312568
\bibitem{RMT} A.Y. Abul-Magd,
Physica A {\bf 361}, 41 (2006)
\bibitem{rapisarda} S. Rizzo and A. Rapisarda,
AIP Conf. Proc. {\bf 742}, 176 (2004) (cond-mat/0406684)
\bibitem{kantz} T. Laubrich, F. Ghasemi, J. Peinke, H. Kantz, arXiv:0811.3337
\bibitem{porpo} A. Porporato, G. Vico, and P.A. Fay,
Geophys. Res. Lett. {\bf 33}, L15402 (2006)
\bibitem{reynolds} A. Reynolds,
Phys. Rev. Lett. {\bf 91}, 084503 (2003)
\bibitem{eco2} H. Aoyama et al.,
arXiv:0805.2792
\bibitem{frank} S.A. Frank and D.E. Smith, Entropy {\bf 12}, 289 (2010)
\bibitem{hase} Y. Hasegawa and M. Arita, arXiv:1004.1452
\bibitem{celia} C. Anteneodo and S.M. Duarte Queiros, J. Stat. Mech. P10023 (2009)
\bibitem{abul-again} A.Y. Abul-Magd, Eur. Phys. J. B {\bf 70}, 39 (2009)
\bibitem{straeten-recent} E. Van der Straeten and C. Beck, arXiv:0911.4816
\bibitem{straetennew} E. Van der Straeten and C. Beck, Phys. Rev. E {\bf 80}, 036108 (2009)
\bibitem{abul-magd3} A.Y. Abul-Magd, G. Akemann, P. Vivo, J. Phys. A Math. Theor. {\bf 42}, 175207 (2009)
%\bibitem{vKa} N.G. van Kampen, {\em Stochastic processes in physics and chemistry},
% North-Holland, London (1982)

%\bibitem{kubo} R. Kubo, M. Toda and N. Hashitsume, {\em Statistical Physics II:
%Nonequilibrium statistical mechanics}, Springer, New York (1995)
%\bibitem{feller} W. Feller, {\em An introduction to probability theory and its applications, Vol. II},
%John Wiley, London (1966)
\bibitem{beck03} C. Beck,
Europhys. Lett. {\bf 64}, 151 (2003)
\bibitem{prl} C. Beck,
Phys. Rev. Lett. {\bf 98}, 064502 (2007)
\bibitem{wilknew} G. Wilk, Z. Wlodarczyk, Eur. Phys. J. A {\bf 40}, 299 (2009)
\bibitem{superscatter} C. Beck, Eur. Phys. J. A {\bf 40}, 267 (2009)
\bibitem{ausloos} M. Ausloos and K. Ivanova, Phys. Rev. E {\bf 68},
046122 (2003)
\bibitem{bouchard} J.-P. Bouchard and M. Potters,
{\em Theory of Financial Risk and Derivative Pricing}, Cambridge
University Press, Cambridge (2003)

\bibitem{abul-magd2} A.Y. Abul-Magd, B. Dietz, T. Friedrich, A. Richer, Phys. Rev. E {\bf 77}, 046202 (2008)
\bibitem{abe-thurner} S. Abe and S. Thurner,
Phys. Rev. E {\bf 72}, 036102 (2005)
\bibitem{queiros} S\'ilvio M. Duarte Queir\'os,  Braz. J. Phys. {\bf 38}, 203 (2008)
\bibitem{briggs} K. Briggs, C. Beck,
Physica A {\bf 378}, 498 (2007)
\bibitem{chen} L. Leon Chen, C. Beck,
Physica A {\bf 387}, 3162 (2008)
\bibitem{abc} S. Abe, C. Beck and G. D. Cohen,
Phys. Rev. E {\bf 76}, 031102 (2007)
\bibitem{crooks} G. E. Crooks,
Phys. Rev. E {\bf 75}, 041119 (2007)
\bibitem{naudts} J. Naudts,
AIP Conference Proceedings {\bf 965}, 84 (2007)


%\bibitem{data_sw} G. S. Lewis and H. L. Swinney, {\it Velocity structure functions, scaling, and
%transitions in high-Reynolds-number Couette-Taylor flow,} Phys. Rev. E {\bf 59}, 5457 (1999)
\bibitem{straeten} E. Van der Straeten, C. Beck,
Phys. Rev. E {\bf 78}, 051101 (2008)
\bibitem{tsallis} C. Tsallis,
J. Stat. Phys. {\bf 52}, 479 (1988)
\bibitem{mendes} C. Tsallis, R.S. Mendes, A.R. Plastino, Physica A
{\bf 261}, 534 (1998)
\bibitem{gentropy} C. Beck, Contemp. Phys. {\bf 50}, 495 (2009)
\bibitem{hau} A.M. Mathai and H.J. Haubold, Physica A {\bf 375}, 110 (2007)

\bibitem{heston} S.L. Heston,
Rev. Fin. Studies {\bf 6}, 327 (1993)
\bibitem{grigolini} S. Bianco, P. Grigolini, P. Paradisi,
cond-mat/0509608
\bibitem{pinton} N. Mordant, P. Metz, O. Michel, and J.-F. Pinton,
Phys. Rev. Lett {\bf 87}, 214501 (2001)

\bibitem{boden-reynolds} A.M. Reynolds, N. Mordant, A.M. Crawford,
and E. Bodenschatz, New Journal of Physics {\bf 7}, 58 (2005)





%\bibitem{jaynes} R. D. Rosenkrantz, {\it E.T. Jaynes: papers on probability, statistics and statistical physics,}
% Kluwer (1989)
%\bibitem{hage} R. Hagedorn, Suppl. Nuovo Cim. {\bf 3}, 147 (1965)
\bibitem{beck00} C. Beck, Physica A {\bf 286}, 164 (2000)
\bibitem{kings} N.E. Mavromatos and S. Sarkar, Phys. Rev. D {\bf 79}, 104015 (2009)
\bibitem{spatio} C. Beck, {\em Spatio-temporal Chaos and Vacuum Fluctuations of
Quantized Fields}, World Scientific, Singapore (2002)
\bibitem{physica-d} C. Beck, Physica D {\bf 171}, 72 (2002)
\end{thebibliography}
\end{document}